\documentclass[twocolumn,showpacs,preprintnumbers,amsmath,amssymb,APS]{revtex4}

\usepackage{dcolumn}
\usepackage{bm}
\usepackage[spanish,english]{babel}
\usepackage{amsfonts}
\usepackage{amssymb}
\usepackage{graphicx}
\usepackage[latin1]{inputenc}
\newcommand{\R}{\mathcal{R}}
\newcommand{\Q}{\mathcal{Q}}

\begin{document}

\title{ {\bf Dynamical Aspects of Generalized Palatini Theories of Gravity}}
\author{Gonzalo J. Olmo}\email{olmo@iem.cfmac.csic.es}
\affiliation{ \footnotesize Instituto de Estructura de la Materia, CSIC, Serrano 121, 28006 Madrid, Spain}

\author{Hèlios Sanchis-Alepuz}
\affiliation{ \footnotesize Departamento de Física Teórica and IFIC, Centro Mixto Universidad de Valencia-CSIC. Facultad de Física, Universidad de Valencia, Burjassot-46100, Valencia, Spain \\ and \\ Fachbereich Theoretische Physik, Institut für Physik, Karl-Franzens-Universität Graz,
Universitätsplatz 5, A-8010 Graz, Austria}

\author{Swapnil Tripathi}
\affiliation{ \footnotesize Physics Department, University of Wisconsin-Milwaukee, Milwaukee, Wisconsin 53201, USA}

\date{June 8$^{th}$, 2009}

\begin{abstract}
We study the field equations of modified theories of gravity in which the lagrangian is a general function of the Ricci scalar and Ricci-squared terms in Palatini formalism. We show that the independent connection can be expressed as the Levi-Cività connection of an auxiliary metric which, in particular cases of interest, is related with the physical metric by means of a disformal transformation. This relation between physical and auxiliary metric boils down to a conformal transformation in the case of $f(R)$ theories. We also show with explicit models that the inclusion of Ricci squared terms in the action can impose upper bounds on the accessible values of pressure and density, which might have important consequences for the early time cosmology and black hole formation scenarios. Our results indicate that the phenomenology of $f(R,R_{\mu\nu}R^{\mu\nu})$ theories is much richer than that of $f(R)$ and $f(R_{\mu\nu}R^{\mu\nu})$ theories and that they also share some similarities with Bekenstein's relativistic theory of MOND. 
\end{abstract}

\pacs{ 04.50.Kd ,98.80.Jk}

\maketitle

\section{Introduction}

The existence of a cosmological constant is a challenging problem for theoretical physics which has gained increasing attention since the discovery of the cosmic speedup \cite{Pad02}. If it is thought as vacuum energy, a naive quantum mechanical analysis raises the question of why it is so small. If quantum field theory in curved space-times is invoked, then the question turns into why it is so large \cite{Wald-Hollands04}. If it is seen as having a quantum gravitational origin, then the dimensionless combination $\Lambda (G\hbar /c^3)\lesssim 10^{-123}$ suggests that this is $123$ orders of magnitude away from the right solution. Interpreting it as a new fundamental constant could imply the existence of new physics in much the same way as the Planck constant did. For these and many other reasons, a strictly constant cosmological term is undesirable. \\ 
\indent Different versions of dark energy sources have been proposed in the literature aimed at relaxing the condition of strict constancy of the so called vacuum energy. A scalar field with very low kinetic energy can closely mimic an effective cosmological constant, though away from the minimum it might exhibit characteristic features that could help distinguish it from a purely constant term. In the arena of modified theories of gravity, there exists a family of theories in which a strictly constant cosmological term arises in certain circumstances. We refer to the so called $f(R)$ theories in Palatini formalism, which are dynamically {\it inequivalent} to their (non-identical) twin brother $f(R)$ in metric formalism. In these theories, the field equations for any lagrangian $f(R)$  become exactly the same as those of General Relativity plus an effective cosmological constant whenever the trace of the energy momentum tensor is constant. When the trace is not constant, one gets modified dynamics induced by the time and/or spatial derivatives of $T$. It is this property of generating an effective cosmological constant that first attracted attention on these theories in relation with the cosmic speedup problem \cite{Palatini}. Any $f(R)$ theory which at low cosmic densities ($T\to 0$) generates a cosmological constant of the right magnitude will produce the desired accelerating effect. However, the $f(R)$ models with infrared corrections considered in the literature so far, besides producing late time cosmic speedup, also lead to catastrophic effects at microscopic scales. Such models induce severe instabilities in atoms due to the strong gravitational backreaction that occurs near the zeros of the atomic wave functions. In those regions the energy density drops virtually to zero and that excites the characteristic infrared scale of the gravitational theory, producing undesired effects which disintegrate the atom \cite{Olmo08a}.\\

The negative features of infrared corrected models, however, are not present in models with just high curvature corrections. Moreover, there is no fundamental reason that prevents an $f(R)$ theory with high curvature corrections from producing an effective cosmological constant compatible with observations. We believe that that line of research has not yet been sufficiently explored. On the other hand, high curvature effects have already been observed in Palatini $f(R)$ theories. In fact, it is possible to find an $f(R)$ Palatini lagrangian which avoids the Big Bang singularity. In \cite{Olmo-Singh09} it was shown that the effective dynamics of Loop Quantum Cosmology (LQC) can be exactly reproduced by an $f(R)$ Palatini theory consisting on an infinite series in $R/R_c$, where $R_c\sim R_{Planck}$ \cite{Olmo-Singh09}. That result is important because it establishes links between an  approach to quantum gravity based on non-perturbative Hamiltonian quantization techniques and a covariant action in Palatini formalism. It is well known that perturbative quantization techniques require the addition of quadratic curvature invariants $R^2, R_{\mu\nu}R^{\mu\nu},\ldots$ (in metric formalism) to the Einstein-Hilbert lagrangian to keep divergences under control. However, it was not known what kind of effective actions, if any, could be associated with the non-perturbative loop quantization until the results of \cite{Olmo-Singh09} were published. \\
\indent Motivated by those results, one may wonder about the role that other curvature invariants besides $R$ might play in the dynamics of Palatini theories. Do $R_{\mu\nu}R^{\mu\nu}$ terms introduce new dynamical effects not present in $f(R)$ theories? Another practical question is how such terms could affect or contribute to the form of the resulting effective cosmological constant in the vacuum limit of the theory. In other words, can we find new insights or mechanisms to obtain a cosmological constant of the right magnitude at low energies? To answer these questions, we are forced to go beyond the simple $f(R)$ models. That is the goal of this paper. \\

Palatini theories with Ricci squared terms have been already studied in the literature. In \cite{ABF04}, the cosmology of $f(R)$ and $f(R_{\mu\nu}R^{\mu\nu})$ theories was considered in some detail. It was found that the scalar $R_{\mu\nu}R^{\mu\nu}$ could be expressed as a function of the trace $T$ of the matter energy-momentum tensor and that new dynamics arises in a way similar to that of $f(R)$ theories, where $R$ can also be expressed as a function of $T$. In \cite{LBM07}, theories of the form $R+ f(R_{\mu\nu}R^{\mu\nu})$ rather than
simply $f(R_{\mu\nu}R^{\mu\nu})$ were considered, and the main focus was on the cosmology at the first-order perturbation level (density perturbation growth). In that work a $3+1$ decomposition was introduced in order to solve the connection equation, and the solution was given at first order in the metric variables. In this work we go a bit farther and consider Palatini theories in which the lagrangian is a generic function of the form $f(R, R_{\mu\nu}R^{\mu\nu})$. We study their field equations and focus on the basic manipulations that will allow us to write the theory in a form suitable for applications. To be precise, we show how to exactly solve the connection equation in terms of the metric and the matter sources (without introducing the $3+1$ decomposition used in \cite{LBM07}), and discuss how the scalars $R$ and $R_{\mu\nu}R^{\mu\nu}$ can be expressed as functions of the density and pressure. We illustrate this point with a family of exactly solvable models and point out that the value of $R_{\mu\nu}R^{\mu\nu}$ in such models is bounded from above, which also sets upper bounds on the energy density and pressure. We comment on the possible consequences that this aspect might have for the early time cosmology of such models. We also find that $R$ and $R_{\mu\nu}R^{\mu\nu}$ are not just functions of the trace $T$, as it happens in the simpler $f(R)$ and $f(R_{\mu\nu}R^{\mu\nu})$ theories. This fact implies that $f(R, R_{\mu\nu}R^{\mu\nu})$ theories have a much richer  phenomenology during the early universe and, in particular, during the radiation dominated era. This is so because in that era, which is characterized by $T=0$, the Palatini corrections of $f(R)$ and $f(R_{\mu\nu}R^{\mu\nu})$ theories boil down to just a cosmological constant, whereas in the $f(R,R_{\mu\nu}R^{\mu\nu})$ case the modified dynamics is more involved. \\

The paper is organized as follows. In section \ref{sec:solving-connections}, we illustrate how to solve for the connection in general $f(R,R_{\mu\nu}R^{\mu\nu})$ theories. We start with the $f(R)$ case to describe the basic steps of the algorithm and then focus on the general $f(R,R_{\mu\nu}R^{\mu\nu})$ case applied to a perfect fluid. Then we write the field equations of the metric in a form suitable for applications and discuss the vacuum limit. In section \ref{sec:R-Q}, we provide a family of models for which $R$ and $R_{\mu\nu}R^{\mu\nu}$ can be explicitly solved in terms of the density $\rho$ and pressure $\Pi$, and show that those quantities must be bounded from above for consistency of the equations.  We conclude with a summary and discussions.

\section{Solving for the connection}\label{sec:solving-connections}

The action that defines a Palatini $f(R,R_{\mu\nu}R^{\mu\nu})$ theory is as follows
\begin{equation}\label{eq:action}
S[g,\Gamma,\psi_m]=\frac{1}{2\kappa^2}\int d^4x \sqrt{-g}f(R,Q) +S_m[g,\psi_m]
\end{equation}
where $g_{\alpha\beta}$ represents the space-time metric, $\Gamma^\alpha_{\beta\gamma}$ is the connection (which is independent of the metric),  $\psi_m$ represents the matter fields, $R=g^{\mu\nu}R_{\mu\nu}$, and $Q=R_{\mu\nu}R^{\mu\nu}$. The Ricci tensor $R_{\mu\nu}={R_{\mu\rho\nu}}^\rho$ is defined in terms of the connection as follows 
\begin{equation}\label{eq:Ricci}
R_{\mu\nu}(\Gamma )=-\partial_{\mu}
\Gamma^{\lambda}_{\lambda\nu}+\partial_{\lambda}
\Gamma^{\lambda}_{\mu\nu}+\Gamma^{\lambda}_{\mu\nu}\Gamma^{\rho}_{\rho\lambda}-\Gamma^{\lambda}_{\nu\rho}\Gamma^{\rho}_{\mu\lambda}
\end{equation}
This definition follows from the relation $[\nabla_\alpha,\nabla_\rho]w_\beta={R_{\alpha\rho\beta}}^\delta w_\delta$ and is valid for any derivative operator $\nabla_\alpha$ \cite{Wald84}.\\
In order to obtain the field equations, we must vary the action with respect to  the various fields present in it. For completeness,  in  Appendix \ref{ap:variation} we detail the process of variation to get the field equations. Variation with respect to metric and connection lead to the following equations
\begin{eqnarray}
f_R R_{\mu\nu}-\frac{f}{2}g_{\mu\nu}+2f_QR_{\mu\alpha}R^\alpha_\nu &=& \kappa^2 T_{\mu\nu}\label{eq:met-varX}\\
\nabla_{\beta}\left[\sqrt{-g}\left(f_R g^{\mu\nu}+2f_Q R^{\mu\nu}\right)\right]&=&0
 \label{eq:con-varX}
\end{eqnarray}
were we have used the short-hand notation $f_R\equiv \partial_R f$, and $f_Q\equiv \partial_Q f$. We now focus on working out a solution for (\ref{eq:con-varX}).\\ 

At first sight, equation (\ref{eq:con-varX}) seems a highly non-trivial equation because of the terms $f_R$ and $f_QR^{\mu\nu}$. In fact, since $R$ and $R^{\mu\nu}$ are functions of the connection and its first derivatives, (\ref{eq:con-varX}) can be seen as a non-linear, second-order equation for the unknown connection. However, there exist algebraic relations between $R$, $R_{\mu\nu}$ and the energy-momentum tensor of the matter that will make the problem easier. To illustrate the general algorithm that solves (\ref{eq:con-varX}), we will first focus on the $f(R)$ case, which is just the case $f_Q=0$.  

\subsection{Solving for the connection in $f(R)$.}
In this case, the connection equation becomes
\begin{equation}\label{eq:con-f_R}
\nabla_{\beta}\left[\sqrt{-g}f_R g^{\mu\nu}\right]=0   \ . 
\end{equation}
From the metric field equation (\ref{eq:met-varX}), we find 
\begin{equation}
f_R R_{\mu\nu}-\frac{f}{2}g_{\mu\nu} = \kappa^2 T_{\mu\nu}\label{eq:met-var1} \ .
\end{equation}
The contraction of this equation with the metric yields
\begin{equation}
f_R R-2f = \kappa^2 T \ .
\end{equation}
This expression is an algebraic equation that generalizes the relation $R=-\kappa^2 T $ of GR to arbitrary lagrangian $f(R)$. This means that given a lagrangian, say $f(R)=R+aR^3$, the above equation can be solved as $R=\R(T)$. In this particular example, the equation is just $-R+aR^3=\kappa^2T$. 
Note that $\R(T)$ is now a function of the matter and, for this reason, $f(\R)$ and $f_R(\R)$ are also functions of the matter. According to this, the connection equation (\ref{eq:con-f_R}) can be seen as a first order equation for the connection that depends on the metric $g_{\alpha\beta}$ and the matter:
\begin{equation}\label{eq:con-f_RT}
\nabla_{\beta}\left[\sqrt{-g}f_R(\R(T)) g^{\mu\nu}\right]=0   \  \ 
\end{equation}
We now discuss how to solve this equation for the connection. In GR, this equation is simply $\nabla_{\beta}\left[\sqrt{-g} g^{\mu\nu}\right]=0$, and the solution  \cite{MTW1973} is given by the Christoffel symbols
\begin{equation}
\Gamma^{\alpha}_{\beta\gamma}=\frac{g^{\alpha\rho}}{2}\left(\partial_\beta g_{\rho\gamma}+\partial_\gamma g_{\rho\beta}-\partial_\rho g_{\beta\gamma}\right)
\end{equation}
In order to find a solution for (\ref{eq:con-f_RT}), we can do the following. We assume that there exists a metric $h_{\mu\nu}$ such that the connection that solves (\ref{eq:con-f_RT}) is the Levi-Cività connection of $h_{\mu\nu}$. This means that the metric $h_{\mu\nu}$ satisfies $\nabla_{\beta}\left[\sqrt{-h} h^{\mu\nu}\right]=0$. In other words, our ansatz satisfies 
\begin{equation}
\sqrt{-h} h^{\mu\nu}=\sqrt{-g}f_{\R} g^{\mu\nu} 
\end{equation}
It is useful to rewrite this equation using matrix notation to suppress indices: $h_{\mu\nu}\to \hat{h}$, $h^{\mu\nu}\to \hat{h}^{-1}$. We then have
\begin{equation}\label{eq:h-matrix-f_R}
\sqrt{-h} \hat{h}^{-1}=\sqrt{-g}f_{\R} \hat{g}^{-1} 
\end{equation}
If we compute the determinant of the left and the right hand sides, we find
\begin{equation}
(\sqrt{-h})^4 h^{-1}=(\sqrt{-g})^4f_\R^4 {g}^{-1} \ \Rightarrow \ h=g f_R^4
\end{equation}
Introducing this result back into (\ref{eq:h-matrix-f_R}), we find $\hat{h}^{-1}=\hat{g}^{-1}/f_\R$, or equivalently, $h^{\mu\nu}=g^{\mu\nu}/f_\R$ and $h_{\mu\nu}=f_\R g_{\mu\nu}$. We thus see that the metric $h_{\mu\nu}$ that defines our independent connection is conformally related to the spacetime metric $g_{\mu\nu}$.\\

At this point, we must make some clarifications about our notation.  
First of all, since we are dealing with two metrics, we can also construct two different Ricci tensors and two different scalar curvatures.  The curvature of the metric $g_{\mu\nu}$ is defined as $R(g)=g^{\mu\nu}R_{\mu\nu}(g)$, where $R_{\mu\nu}(g)$ is the Ricci tensor constructed from the Levi-Cività connection of $g_{\mu\nu}$. Analogously, we can define the scalar curvature of the metric $h_{\mu\nu}$ as the contraction $R(h)=h^{\mu\nu}R_{\mu\nu}(h)$, which is equivalent to $h^{\mu\nu}R_{\mu\nu}(\Gamma)$, where $\Gamma$ is the Levi-Cività connection of $h_{\mu\nu}$. Now, in the above equations we have been dealing with something different from $R(g)$ and $R(h)$. What we called $R$ or $\R(T)$ is a hybrid object, it is the contraction  $g^{\mu\nu}R_{\mu\nu}(\Gamma)$, which is defined in terms of the independent connection via $R_{\mu\nu}(\Gamma)=R_{\mu\nu}(h)$ and the spacetime metric $g^{\mu\nu}$. In all our previous formulas, we dealt with the object $\R$, which is a function of the matter via the trace $T$. Whenever we write $\R$ it should be understood that it is a function of the matter. In any other case, we will write either $R(g)$ or $R(h)$.\\

For completeness, let us now focus on the equation for the metric. In order to write a second-order equation for the metric, we must use the solution that we just found for the connection and introduce the result in (\ref{eq:met-var1}), which is written in terms of $R_{\mu\nu}(h)$ and $g_{\mu\nu}$. Since $\hat{h}$ and $\hat{g}$ are conformally related, it is straightforward to relate $R_{\mu\nu}(h)$ with $R_{\mu\nu}(g)$ (see Appendix D of \cite{Wald84}). The result is that (\ref{eq:met-var1}) becomes
\begin{eqnarray}\label{eq:neweinstein}
G_{\mu \nu}(g) &=& \frac{\kappa^2}{f_R} T_{\mu \nu} - \frac{\R f_R - f}{2 f_R} g_{\mu \nu} + \frac{1}{f_R}\left(\nabla_\mu \nabla_\nu f_R - g_{\mu \nu} \Box f_R\right) \nonumber\\ 
&& - \frac{3}{2 f_R^2} \left(\partial_\mu f_R \partial_\nu f_R - \frac{1}{2} g_{\mu \nu} (\partial f_R)^2\right) 
\end{eqnarray}
where now $\nabla_\mu$ is the usual covariant derivative of $g_{\mu\nu}$. 
The lesson to learn here is that the right hand side of (\ref{eq:neweinstein}) behaves like a modified energy-momentum tensor in which the trace $T$ plays a role non-existing in GR via the terms $\R,f(\R)$ and the various derivatives of $f_R(\R)$. When $T$ is constant, all $\partial_\mu f$ terms vanish and the equations boild down to those of GR plus an effective cosmological constant given by $\frac{\R f_R - f}{2 f_R}$ evaluated at constant $T$.

\subsection{Solving for the connection in general $f(R,Q)$}

We now face the problem of solving the constraint equation (\ref{eq:con-varX}). 
In this case, the metric variation leads to 
\begin{equation}
f_R R_{\mu\nu}-\frac{f}{2}g_{\mu\nu}+2f_QR_{\mu\alpha}R^\alpha_\nu = \kappa^2 T_{\mu\nu}\label{eq:met-varRQ}
\end{equation}
In order to solve (\ref{eq:con-varX}) we will follow the same steps as above but adding the complexity inherent to the new problem. \\
\begin{itemize}
\item {\bf Step 1.} We need to find an algebraic relation between $R$ and $R_{\mu\nu}$ with the matter sources. This will allow us to re-interpret (\ref{eq:con-varX}) as a first order equation for the connection. To proceed, we first define the matrix $\hat{P}$, whose components are ${P_\mu}^\nu\equiv R_{\mu\alpha}g^{\alpha\nu}$, which allows us to express (\ref{eq:met-varRQ}) as
\begin{equation}
f_R {P_\mu}^\nu-\frac{f}{2}{\delta_\mu}^\nu+2f_Q{P_\mu}^\alpha {P_\alpha}^\nu= \kappa^2 {T_\mu}^\nu\label{eq:met-varRQ1} \ .
\end{equation}
In matrix notation, this equation reads 
\begin{equation}
2f_Q\hat{P}^2+f_R \hat{P}-\frac{f}{2}\hat{I} = \kappa^2 \hat{T} \label{eq:met-varRQ2} \ ,
\end{equation}
where $\hat{T}$ is the matrix representation of ${T_\mu}^\nu$. Note that $R$ and $Q$ are the trace of $\hat{P}$ and the trace of $\hat{P}^2$ respectively. Solving this equation will thus lead to a relation of the form $\hat{P}=\hat{P}(\hat{T})$, which is analogous to the solution $\R(T)$ of the $f(R)$ case. For the moment we will not care about the particular form of this solution, which depends on the particular model chosen, and will just assume that it exists. This is what we need to reinterpret (\ref{eq:con-varX}) as a first order equation for the connection. The solution should thus depend on the metric $g_{\mu\nu}$ and the $T_{\mu\nu}$ of the matter. 
\item {\bf Step 2.} We now propose an ansatz of the same form as above. We look for a metric $\hat{h}$ such that $\nabla_{\beta}\left[\sqrt{-h} h^{\mu\nu}\right]=0$. This guarantees that the connection will be the Levi-Cività connection of $\hat{h}$. Using matrix notation, we have
\begin{equation}
\sqrt{-h}\hat{h}^{-1}=\sqrt{-g}\hat{g}^{-1}\left(f_R\hat{I}+2f_Q\hat{P}\right) \ .
\end{equation}
We now compute the determinant of the left and the right hand sides, which give $h=g \det\left(f_R\hat{I}+2f_Q\hat{P}\right)$. Once we know the explicit expression for $\hat{P}$ we will be able to compute this determinant. In any case, we have the formal expression
\begin{equation}
\hat{h}^{-1}=\frac{\hat{g}^{-1}\hat\Sigma}{\sqrt{\det\hat\Sigma}} \ ,
\end{equation}
where we have defined $\hat\Sigma=\left(f_R\hat{I}+2f_Q\hat{P}\right)$. Taking the inverse of the above matrix, we find 
\begin{equation}
\hat{h}=\left(\sqrt{\det\hat\Sigma}\right)\hat\Sigma^{-1}\hat{g} \ .
\end{equation}
\end{itemize}
We have thus shown that the connection of $f(R,Q)$ theories can be explicitly solved in terms of the physical metric $g_{\mu\nu}$ and the matter sources. To proceed further, we need to consider particular choices for $T_{\mu\nu}$ and/or particular models. This is the task of the next sections.

\subsection{$f(R,Q)$ with a perfect fluid}

We now consider the case of matter described by ${T_\mu}^\nu=\Pi{\delta_\mu}^\nu+(\rho+\Pi)u_\mu u^\nu$, where $\Pi$ represents pressure to avoid missunderstanding with the matrix $\hat{P}$. The first thing that we need to do is to compute the matrix $\hat\Sigma$ and its inverse, because that will allow us to compute $\hat{h}$ and $\hat{h}^{-1}$. To find $\hat\Sigma$, we first need to find $\hat P$, which is a solution of (\ref{eq:met-varRQ2}). That equation can be rewritten as follows:
\begin{equation}
2f_Q\left(\hat P+\frac{f_R}{4f_Q}\hat I\right)^2=\left(\kappa^2\Pi+\frac{f}{2}+\frac{f_R^2}{8f_Q}\right)\hat I+\kappa^2(\rho+\Pi)u_\mu u^\mu\label{eq:met-varRQ3} \ ,
\end{equation} 
which can formally be denoted as
\begin{equation}
2f_Q\hat M^2=\alpha\hat I+\beta u_\mu u^\nu \label{eq:M_1} \ ,
\end{equation}
so that we can look for a solution of the form
\begin{equation}
\sqrt{2f_Q}\hat M=\lambda I+\sigma u_\mu u^\nu \label{eq:M_2} \ ,
\end{equation}
where $f_Q>0$ has been assumed. We can now take the square of this matrix to find the relation between $(\lambda,\sigma)$ and $(\alpha,\beta)$:
\begin{equation}\label{eq:lambda-sigma}
\lambda^2=\alpha  \ , \ \sigma=\frac{(-\lambda\pm\sqrt{\lambda^2+\beta n_u})}{n_u} \ ,
\end{equation}
where $n_u\equiv u_\mu u^\mu$ is the norm of the vector $u_\mu$, which for the perfect fluid is just $n_u=-1$.  If $f_Q<0$, then the same manipulations hold up to the redefinitions $f_Q\to -|f_Q|$, $\alpha\to -\alpha$, and $\beta\to -\beta$. 
The matrix $\hat \Sigma$ can then be written as
\begin{equation}\label{eq:Sigma}
{\Sigma_\mu}^\nu=\Lambda_1\delta_\mu^\nu+\Lambda_2u_\mu u^\nu \ ,
\end{equation}
where we have defined 
\begin{eqnarray}
\Lambda_1&=& \sqrt{2f_Q}\lambda+\frac{f_R}{2}\\
\Lambda_2&=& \sqrt{2f_Q}\sigma
\end{eqnarray}
The determinant of a matrix of the form (\ref{eq:Sigma}) can be computed straightforwardly using the definition $\det \hat M= \epsilon_{abcd}{M_0}^a{M_1}^b{M_2}^c{M_3}^d$ and leads to 
\begin{equation}
\det{\hat\Sigma}= \Lambda_1^3 (\Lambda_1+n_u \Lambda_2) \ .
\end{equation}
The inverse ${(\Sigma^{-1})_\mu}^\nu$ has the form
\begin{equation}
{(\Sigma^{-1})_\mu}^\nu=\frac{1}{\Lambda_1}\delta_\mu^\nu-\frac{\Lambda_2}{\Lambda_1}\frac{1}{(\Lambda_1+\Lambda_2n_u)}u_\mu u^\nu
\end{equation}
We are thus ready to express the metric in terms of known quantities. 
\begin{eqnarray}\label{eq:hmn1}
h_{\mu\nu}&=&\Omega\left[g_{\mu\nu}-\frac{\sqrt{2f_Q}\sigma}{\left[\sqrt{2f_Q}(\lambda+n_u \sigma)+\frac{f_R}{2}\right]} u_\mu u_\nu\right]\\
h^{\mu\nu}&=&\frac{1}{\Omega}\left[g^{\mu\nu}+\frac{\sqrt{2f_Q}\sigma}{\left[\sqrt{2f_Q}\lambda+\frac{f_R}{2}\right]} u^\mu u^\nu\right]\label{eq:hmn2}
\end{eqnarray}
where $\Omega=\sqrt{\det{\hat\Sigma}}/\Lambda_1$. It is worth noting that in the limit $f_Q\to 0$ we find $\Omega\to f_R$ and $\sigma\to 0$ when the plus sign in front of the square root of (\ref{eq:lambda-sigma}) is taken, which recovers the equations of $f(R)$ theories. A relation between two metrics of the form found in (\ref{eq:hmn1}) and (\ref{eq:hmn2}), namely $h_{\mu\nu}=\Omega_1 g_{\mu\nu}-\Omega_2 u_\mu u_\nu$, involving a vector field and two independent functions $\Omega_1$ and $\Omega_2$, is known in the literature as ``disformal transformation''. A relation of this form between two metrics  has already been studied in inflationary models characterized by non-minimal couplings between curvature and a scalar field 
\cite{CLS2000}, and naturally arises in Bekenstein's relativistic theory of MOND \cite{Bekenstein04} and in previous versions of it. In the MOND theory, the vector $u_\mu$ is an independent dynamical vector field and the functions in front of it and in front of $g_{\mu\nu}$ depend on another dynamical scalar field. In the theory described here, on the contrary, the metric tensor is the only dynamical field of the gravitational sector.

\section{Field equations and $\Lambda_{eff}$}

To find the field equations for the metric in a way that reminds those of GR, we can take (\ref{eq:met-varRQ1}) and put it as
\begin{equation}\label{eq:Ricci-1}
{P_\mu}^\alpha{\Sigma_\alpha}^\nu=\kappa^2{T_\mu}^\nu+\frac{f}{2}{\delta_\mu}^\nu \ .
\end{equation}
Using the inverse of $\hat\Sigma$ and the definition of ${P_\mu}^\alpha=R_{\mu\nu}g^{\nu\alpha}$, we can express these equations as 
\begin{equation}\label{eq:Ricci-2}
R_{\mu\nu}(\Gamma)=(\kappa^2{T_\mu}^\alpha+\frac{f}{2}{\delta_\mu}^\alpha){(\Sigma^{-1})_\alpha}^\beta g_{\beta\nu}\equiv \tau_{\mu\nu} \ .
\end{equation}
In the case of a perfect fluid, $n_u=-1$, the right hand side of this equation becomes
\begin{equation}\label{eq:Tmn-pf}
\tau_{\mu\nu}=\frac{\left(\frac{f}{2}+\kappa^2\Pi\right)}{\Lambda_1}g_{\mu\nu}+\frac{\Lambda_1\kappa^2(\rho+\Pi)-\Lambda_2 \left(\frac{f}{2}+\kappa^2\Pi\right)}{\Lambda_1(\Lambda_1-\Lambda_2)}u_\mu u_\nu
\end{equation}
The field equations for the metric $g_{\mu\nu}$ can now be obtained from (\ref{eq:Ricci-2}) by expressing the connection $\Gamma(h)$ in terms of $g_{\mu\nu}$ and the matter using (\ref{eq:hmn1}) and (\ref{eq:hmn2}). In the case of $f(R)$ theories, this process leads to (\ref{eq:neweinstein}). (Recall that the $f(R)$ field equations are recovered in the limit $f_Q\to 0$ when the plus sign in front of the square root of (\ref{eq:lambda-sigma}) is taken.) In the case of $f(R,Q)$ theories, we find it more convenient to work directly with (\ref{eq:Ricci-2}). Examples of applications will be given elsewhere \cite{OST-ip}.\\ 

Let us now consider the effective cosmological constant $\Lambda_{eff}$ that follows from a generic $f(R,Q)$ theory in vacuum. 
With the same choice of sign that recovers the right $f(R)$ limit, (\ref{eq:Ricci-2}) in vacuum becomes
\begin{equation}\label{eq:Ricci-3}
{R_\mu}^\nu(\Gamma)_{vac}=\left.\frac{f}{2\Lambda_1}\right|_{\rho,\Pi=0}{\delta_\mu}^\nu \ .
\end{equation}
Since in vacuum the relation between $h_{\mu\nu}$ and $g_{\mu\nu}$ is just a constant conformal factor, $h_{\mu\nu}=\left.\Omega\right|_{{\rho,\Pi=0}}g_{\mu\nu}$ (with $\left.\Omega\right|_{{\rho,\Pi=0}}=\left.\Lambda_1\right|_{{\rho,\Pi=0}}$),  it follows that $R_{\mu\nu}(\Gamma)_{vac}=R_{\mu\nu}(g)_{vac}$, which implies that 
\begin{equation}\label{eq:Ricci-vac}
R_{\mu\nu}(g)_{vac}=\left.\frac{f}{2\Lambda_1}\right|_{{\rho,\Pi=0}} g_{\mu\nu}=\Lambda_{eff}g_{\mu\nu} .
\end{equation}
This equation indicates that $\Lambda_{eff}$ is the result of evaluating the lagrangian $f$ in vacuum, up to a factor $1/2\Lambda_1$. Now, from the trace of (\ref{eq:met-varRQ2}) we see that in vacuum $f/2=(R f_R+2Q f_Q)/4$. Since we assume lagrangians $f$ which recover GR at low curvatures, which implies that $f_R\to 1$ in vacuum, it follows that to get a non-zero $\Lambda_{eff}$ we must have $R_{vac}\neq 0$, or $Q f_Q\neq 0$ or a combination of both. In any case, we see that the inclusion of $Q$-dependent terms in the lagrangian provides new mechanisms besides those present in $f(R)$ theories to generate a non-zero cosmological constant.

\section{Solving for $R(\rho,\Pi)$ and $Q(\rho,\Pi)$.}\label{sec:R-Q}

In order to study physical predictions of the theories considered in this work, we must specify particular models. Once a model is chosen, the problem of finding $R=\R(\rho,\Pi)$ and $Q=\Q(\rho,\Pi)$ is difficult in general because it can involve non-linear equations with multiple solutions. In principle, one only needs two independent equations relating $R,Q$ with $\rho,\Pi$ but it will not be obvious always how to construct two such equations to make the problem analytically tractable. Here we will impose some restrictions on the lagrangian to simplify and illustrate the problem. We will consider $f(R,Q)$ functions of the form $f(R,Q)=\tilde f(R)+\frac{R_{\mu\nu}R^{\mu\nu}}{R_P}$, where $R_P$ is a constant of the order of the Planck curvature. The reason for this choice becomes apparent when we take the trace of equation (\ref{eq:met-varRQ2})
\begin{equation}
2Qf_Q +Rf_R-2f=\kappa^2T 
\end{equation}  
This equation relates $R$ and $Q$ with $T$, but in the particular case of  $f(R,Q)=\tilde f(R)+R_{\mu\nu}R^{\mu\nu}/R_P$, it becomes
\begin{equation}
R\tilde f_R-2\tilde f=\kappa^2T 
\end{equation}  
which is exactly the same expression as in $f(R)$ theories and implies that $R=\R(T)$. 
There still remains to obtain $Q=Q(\rho,\Pi)$. This function can be found taking the trace of (\ref{eq:M_2})  
\begin{equation}
\sqrt{2f_Q}\left(R+\frac{f_R}{f_Q}\right)=3\lambda+
\sqrt{\lambda^2-\beta} \ ,
\end{equation}
which can be cast as
\begin{equation}
\left[\sqrt{2f_Q}\left(R+\frac{f_R}{f_Q}\right)-3\lambda\right]^2=
\lambda^2-\beta
\end{equation}
After a bit of algebra we find that 
\begin{equation}\label{eq:lambda1}
\lambda= \frac{\sqrt{2f_Q}}{8}\left[3\left(R+\frac{f_R}{f_Q}\right)\pm\sqrt{\left(R+\frac{f_R}{f_Q}\right)^2-\frac{4\beta}{f_Q}}\right]
\end{equation}
For our choice of $f(R,Q)$ we see that the $f_Q$ term appearing on the right hand side of (\ref{eq:lambda1}) is just the constant $f_Q=1/R_P$. We can then use the definition of $\lambda^2$ [see (\ref{eq:met-varRQ3}), (\ref{eq:M_1}), and (\ref{eq:M_2})] to solve for $Q$ as a function of $\R(T),\rho$, and $\Pi$. The solution is 
\begin{eqnarray}\label{eq:lambda2}
\frac{Q}{2R_P}&=&-\left(\kappa^2\Pi+\frac{\tilde f}{2}+\frac{R_P}{8}\tilde f_R^2\right)+\frac{{R_P}}{32}\left[3\left(\frac{R}{R_P}+\tilde f_R\right)\right. \nonumber \\ &\pm&\left.\sqrt{\left(\frac{R}{R_P}+\tilde f_R\right)^2-\frac{4\kappa^2(\rho+\Pi)}{R_P}}\right]^2
\end{eqnarray}
The correct limit when $R_P\to\infty$ is obtained for the {\bf minus sign} in front of the square root. Had we chosen $R_P$ negative, i.e., $R_P\to -R_P$, (\ref{eq:lambda2}) would still hold but the positive sign in front of the square root should be needed to obtain the right limit. This can be easily verified for particular choices of $f$.\\
\indent It is remarkable that (\ref{eq:lambda2}) with $R_P>0$ implies that $Q$ is bounded in the real line. Consistency requires that the term under the square root be non negative, i.e.,  
\begin{equation}\label{eq:ineq}
\left(\frac{R}{R_P}+\tilde f_R\right)^2-\frac{4\kappa^2(\rho+\Pi)}{R_P}\ge 0
\end{equation}
which places bounds on the accessible values of $\rho$ and $\Pi$.  To illustrate this point, let us consider the family of lagrangians $f(R,Q)=R+aR^2/R_P+Q/R_P$, for which $R$ turns out to go exactly like in GR, $\R=-\kappa^2T$. If in this example we choose $a=-1/2$, i.e., $\tilde{f}(R)=R-R^2/(2R_P)$, then (\ref{eq:lambda2}) becomes
\begin{eqnarray}\label{eq:Q-1/2}
Q&=&\frac{3R_P^2}{8}\left[1-\frac{2\kappa^2(\rho+\Pi)}{R_P}+\frac{2\kappa^4(\rho-3\Pi)^2}{3R_P^2}\right.\nonumber \\ &-&\left.\sqrt{1-\frac{4\kappa^2(\rho+\Pi)}{R_P}}\right] \ .
\end{eqnarray}
At low energies, this expression recovers the GR limit 
\begin{equation}
Q\approx \left(3 P^2+\rho ^2\right)+\frac{3 (P+\rho )^3}{2 R_P}+\frac{15 (P+\rho )^4}{4 {R_P}^2}+\ldots
\end{equation}
However, positivity of the argument in the square root of (\ref{eq:Q-1/2}) implies that at high energies 
\begin{equation}
\kappa^2(\rho+\Pi)\leq \frac{R_P}{4} \ ,
\end{equation}
which clearly shows that the combination $\rho+\Pi$ is bounded from above. 
Taking an equation of state of the form $\Pi=w\rho$, we find that the maximum value of $Q$ occurs at $\kappa^2\rho_{max}=R_P/(4+4w)$ and is given by 
\begin{equation}
Q_{max}=\frac{3R_P^2}{16}\left[1+\frac{(1-3w)^2}{12(1+w)^2}\right]
\end{equation}

It is important to note that the upper bound on $\rho$ and $\Pi$ is a consequence of the field equations, which have been used to obtain (\ref{eq:lambda2}), and is independent of the symmetries or equation of state that our particular problem might have. As a consequence, we may expect to observe important new effects in critical scenarios such as in stellar collapse processes or in the very early Universe, where singularities are unavoidable in the context of GR. 
We would like to remark that, within the metric formalism, theories with bounded curvature scalars were proposed time ago in \cite{Bran-Muk92} to construct non-singular isotropic universes. In such constructions, the Big Bang singularity is cured by the self-interactions introduced by two scalar fields, whose role was to place bounds on the upper limits accessible to certain curvature scalars. In the theory discussed here, no new degrees of freedom have been introduced. It is the Palatini dynamics that constraints the physical range of $\rho$ and $\Pi$ and, therefore, suggests that all curvature invariants will be bounded. This contrasts with the generic behavior of the models $R+(aR_{\mu\nu}R^{\mu\nu}+bR^2)/R_P$ in metric formalism, where all solutions which at late times recover a FLRW Universe start with a Big Bang singularity and, therefore, with divergent scalars $R$ and $Q$ \cite{MuSchSta85-88}.

\section{Summary and conclusions}

In this work we have studied the field equations of Palatini theories of gravity in which the lagrangian is a function of the form $f(R,R_{\mu\nu}R^{\mu\nu})$. We have shown that the independent connection can be expressed as the Levi-Cività connection of an auxiliary metric which is related to the physical metric $g_{\mu\nu}$ and the energy-momentum tensor by means of a non-standard transformation, which becomes disformal when matter is described as a perfect fluid and boils down to conformal when the $R_{\mu\nu}R^{\mu\nu}$ dependence disappears. The emergence of two metrics related by a disformal transformation, a basic requirement of relativistic MOND theories to properly account for gravitational lensing, could make these theories interesting for the consideration of dark-matter-related problems. \\
\indent We have also shown that in Palatini $f(R,R_{\mu\nu}R^{\mu\nu})$ theories the scalars $R$ and $R_{\mu\nu}R^{\mu\nu}$ can in general be written as functions of $\rho$ and $\Pi$ but not necessarily via the trace $T$. As a result, the phenomenology of these theories is much richer than that of the individual $f(R)$ or $f(R_{\mu\nu}R^{\mu\nu})$ theories. Furthermore, for some simple models, we have shown explicitly that the scalar $R_{\mu\nu}R^{\mu\nu}$ sets bounds on the physically accessible range of $\rho$ and $\Pi$, which suggests that scenarios such as the very early Universe and the last stages of stellar collapse could be seriously affected by the new dynamics possibly leading to singularity resolution. \\
\indent The results obtained in this work open new avenues of research in the context of the very early Universe, the radiation dominated epoch, and the accelerating Universe, with new mechanisms to generate an effective cosmological constant, within a framework that does not introduce new degrees of freedom and, therefore, is closer to GR than other type of modified theories of gravity or dark energy models. 

\acknowledgements

S.T. thanks K. Taniguchi and H. Markakis for useful comments and discussions. H.S. thanks J.Navarro-Salas for his continuous support and advice. G.J.O. thanks MICINN for a Juan de la Cierva contract, the Physics Department of the University of Wisconsin-Milwaukee, and the Departamento de Física Teórica \& IFIC  of the University of Valencia - CSIC  for their hospitality during the elaboration of this work. The research of G.J.O. has also been partially supported by grant FIS2005-05736-C03-03. S.T. has been supported by NSF-0701817.

\appendix

\section{Variation of the action}\label{ap:variation}

The variation of the action (\ref{eq:action}) can be expressed as
\begin{equation}
\delta S=\frac{1}{2\kappa^2}\int d^4x \sqrt{-g}\left[-\frac{f}{2}g_{\mu\nu}\delta g^{\mu\nu}+\delta f(R,Q)\right] \ ,
\end{equation}
where $\delta f(R,Q)$ represents
\begin{equation}
\delta f(R,Q)=f_R \delta R+f_Q \delta Q
\end{equation}
and the subindex in $f_R$ and $f_Q$ denotes partial derivative.
Since $R=g^{\mu\nu}R_{\mu\nu}$ and $Q=g^{\mu\alpha}g^{\nu\beta}R_{\mu\nu}R_{\alpha\beta}$, it is easy to see that 
\begin{eqnarray}
\delta R&=&\delta(g^{\mu\nu}R_{\mu\nu}) = R_{\mu\nu}\delta g^{\mu\nu}+g^{\mu\nu}\delta R_{\mu\nu} \\
\delta Q&=&\delta (g^{\mu\nu}g^{\alpha\beta}R_{\mu\alpha}R_{\nu\beta}) \nonumber\\&=&2R_{\mu\alpha}R^\alpha_\nu \delta g^{\mu\nu}+2R^{\mu\nu}\delta R_{\mu\nu}
\end{eqnarray}
Inserting these results in $\delta S$ we find
\begin{eqnarray}
\delta S&=&\frac{1}{2\kappa^2}\int d^4x \sqrt{-g}\Big[\Big(f_R R_{\mu\nu}-\frac{f}{2}g_{\mu\nu} + 2f_QR_{\mu\alpha}R^\alpha_\nu\big)\delta g^{\mu\nu} \nonumber\\& &+ \left(f_R g^{\mu\nu}+2f_Q R^{\mu\nu}\right)\delta R_{\mu\nu}\big]
\end{eqnarray}
The next step requires to express $\delta R_{\mu\nu}$ in terms of $\delta \Gamma^\alpha_{\mu\nu}$. This can be done using the so called Palatini identity 
\begin{equation}
\delta R_{\mu\nu}= -\nabla_\mu \left(\delta \Gamma^\lambda_{\lambda\nu}\right)+\nabla_\lambda \left(\delta\Gamma^\lambda_{\mu\nu}\right)
\end{equation}
We now manipulate the $\delta R_{\mu\nu}$ term. That contribution is of the form 
\begin{equation}
M=\int d^4 x\sqrt{-g}\Lambda^{\mu\nu}\delta R_{\mu\nu}
\end{equation}
where, in our case, $\Lambda^{\mu\nu}\equiv\left(f_R g^{\mu\nu}+2f_Q R^{\mu\nu}\right)$. 
Using the Palatini identity, we get
\begin{equation}
M=\int d^4 x\sqrt{-g}\Lambda^{\mu\nu}\left[-\nabla_\mu \left(\delta \Gamma^\lambda_{\lambda\nu}\right)+\nabla_\lambda \left(\delta\Gamma^\lambda_{\mu\nu}\right)\right]
\end{equation}
Using integration by parts and rearranging indices, we find
\begin{eqnarray}
M=\int d^4x\Big\{\nabla_\lambda\left[\sqrt{-g}\left(\Lambda^{\mu\nu}\delta\Gamma^\lambda_{\mu\nu}-\Lambda^{\lambda\nu}\delta\Gamma^\rho_{\rho\nu}\right)\right]+\nonumber\\ \nabla_\mu\big[\sqrt{-g}\big(\Lambda^{\mu\nu}\delta^\lambda_\beta-\delta^\mu_\beta\Lambda^{\lambda\nu}\big)\big]\delta\Gamma^\beta_{\lambda\nu} \Big\}
\end{eqnarray}
The first term in brackets is a total derivative and can be discarded. The second term is the one we need.
Putting this back into $\delta S$ we end up with
\begin{eqnarray}
\delta S &= & \frac{1}{2\kappa^2}  \int  d^4x \sqrt{-g}\Big\{ \big(f_R R_{\mu\nu}-\frac{f}{2}g_{\mu\nu}+2f_QR_{\mu\alpha}R^\alpha_\nu\big)\delta g^{\mu\nu} \nonumber\\ &  &+\nabla_\mu\left[\sqrt{-g}\big(\Lambda^{\mu\nu}\delta^\lambda_\beta-\delta^\mu_\beta\Lambda^{\lambda\nu}\big)\right]\delta\Gamma^\beta_{\lambda\nu}\Big\}
\end{eqnarray}
Knowing that the matter action gives $\delta S_m=-\frac{1}{2}\int d^4x\sqrt{-g}T_{\mu\nu}\delta g^{\mu\nu}$, the field equations can be written as follows
\begin{eqnarray}
f_R R_{\mu\nu}-\frac{f}{2}g_{\mu\nu}+2f_QR_{\mu\alpha}R^\alpha_\nu &=& \kappa^2 T_{\mu\nu}\label{eq:met-var0}\\
\nabla_\mu\left[\sqrt{-g}\left(\Lambda^{\mu\nu}\delta^\lambda_\beta-\delta^\mu_\beta\Lambda^{\lambda\nu}\right)\right]&=&0 \label{eq:con-var0}
\end{eqnarray}
Note that the second equation is equated to zero because matter is not coupled to the independent connection ($\delta\Gamma$ does not appear in $\delta S_m$). Equation (\ref{eq:con-var0}) can be further simplified if one notices that when $\lambda=\beta$ then the equation is identically zero. Taking $\lambda\neq \beta$ then it boils down to $\nabla_{\beta}\left[\sqrt{-g}\Lambda^{\lambda\nu}\right]=0$, which is explicitly given by  
\begin{equation}\label{eq:constraint}
\nabla_{\beta}\left[\sqrt{-g}\left(f_R g^{\mu\nu}+2f_Q R^{\mu\nu}\right)\right]=0
\end{equation}
It is straightforward to verify that when $f=f(R)$, no explicit dependence on $Q$, all the above equations reduce to the case of Palatini $f(R)$.

\end{document}